\documentclass[aps,prb,preprint,11pt,showkeys,nobibnotes,floatfix]{revtex4}

\usepackage{amsmath,graphicx}

\begin{document}

%%%%%%%%%%%%%%%%%%%%%%%%%%%%%%%%%%%%%%%%%%%%%%%%%%%%%%%%%%%%%%%%%%%%%%%%
%                          Title and Authors
%%%%%%%%%%%%%%%%%%%%%%%%%%%%%%%%%%%%%%%%%%%%%%%%%%%%%%%%%%%%%%%%%%%%%%%%

\title{Ultrafast Optical Manipulation of Ferromagnetic Order in
InMnAs/GaSb}

\author{J. Wang}
\author{G. A. Khodaparast}
\author{J. Kono}
\email[Please send all correspondence to: Prof. Junichiro Kono,
Rice University, ECE Dept., MS-366, P.O. Box 1892,
Houston, TX 77251-1892, U.S.A.  Phone: +1-713-348-2209.
E-mail:]{kono@rice.edu}
\affiliation{Department of Electrical and Computer Engineering,
Rice Quantum Institute, and Center for Nanoscale Science
and Technology, Rice University, Houston, Texas 77005, U.S.A.}

\author{T. Slupinski}
\affiliation{Institute of Experimental Physics, Warsaw University,
Hoza 69, 00-681 Warsaw, Poland}

\author{A. Oiwa}
\author{H. Munekata}
\affiliation{Imaging Science and Engineering Laboratory,
Tokyo Institute of Technology, Yokohama, Kanagawa 226-8503, Japan}

\date{\today}

%%%%%%%%%%%%%%%%%%%%%%%%%%%%%%%%%%%%%%%%%%%%%%%%%%%%%%%%%%%%%%%%%%%%%%%%
%                            Abstract
%%%%%%%%%%%%%%%%%%%%%%%%%%%%%%%%%%%%%%%%%%%%%%%%%%%%%%%%%%%%%%%%%%%%%%%%

\begin{abstract}

We have performed a two-color time-resolved magneto-optical Kerr effect
(MOKE) study of a ferromagnetic InMnAs/GaSb heterostructure.
We observed ultrafast photo-induced changes in the MOKE signal induced
by a large density of spin-polarized transient carriers created only within
the InMnAs layer using intense 140 fs mid-infrared pulses.  Our data clearly
demonstrates that magnetic properties, e.g., remanence and coercivity,
can be strongly modified.  The dependence of these
changes on the time delay, pump polarization, pump intensity, and
sample temperature is discussed.

\end{abstract}

\maketitle

%%%%%%%%%%%%%%%%%%%%%%%%%%%%%%%%%%%%%%%%%%%%%%%%%%%%%%%%%%%%%%%%%%%%%%%%
%                           Body of Paper
%%%%%%%%%%%%%%%%%%%%%%%%%%%%%%%%%%%%%%%%%%%%%%%%%%%%%%%%%%%%%%%%%%%%%%%%

%-----------------------------------------------------------------------
\section{Introduction}
%-----------------------------------------------------------------------

Recently, there has been significant interest in ultrafast spin dynamics in
ferromagnets, both from scientific and technological viewpoints.\cite{zhang}
Itinerant ferromagnets such as nickel, cobalt, iron, and CoPt$_3$ have been
studied extensively using ultrafast optical and magneto-optical
spectroscopies, exhibiting an array of new phenomena.\cite{beaurepaire,
hohlfeld,scholl,aeschlimann,ju,beaurepaire1,gudde,koopmans,guidoni}  In
particular, the discovery of {\em ultrafast demagnetization}\cite{beaurepaire}
suggested a novel ultrafast scheme for writing data in magneto-optical
recording applications.
At the same time, exactly how a laser pulse can effectively change the
magnetic spin moment in an ultrafast manner is an open question, motivating
intensive experimental and theoretical investigations.\cite{gudde,koopmans,
guidoni,zhang1}  In extreme cases, intense laser pulses were shown to
increase the electron temperature to even above the Curie temperature,
driving a ferromagnetic to paramagnetic phase transition in the
femtosecond time scale.\cite{beaurepaire1}

III-V ferromagnetic semiconductors such as InMnAs\cite{ohno,munekata} and
GaMnAs\cite{ohno1} can add new dimensions to this problem.  The
{\em carrier-induced} nature of ferromagnetism in these semiconductors,
whose microscopic origin is a matter of controversy,\cite{theory} has
paved a natural path to electrical\cite{ohno2} and optical\cite{koshihara}
control of magnetic order.  Since ultrashort laser pulses can create a large
density of {\em transient carriers} in semiconductors, in addition to
heating the electron system as in the case of ferromagnetic metals,
one can anticipate significant modifications in the exchange interaction
between Mn ions.  In addition, unlike metals, pumping semiconductors with
circularly-polarized light results in spin-coherent carriers, which
should not only lead to their own contribution to the net magnetization,
but also be able to enhance or reduce the magnetization due to the Mn spins
via $s-d$ and $p-d$ exchange interactions, depending on the relative
orientations of the carrier spins and localized Mn spins.

Here we report results of an ultrafast optical study of spin/magnetization
dynamics in a ferromagnetic InMnAs/GaSb heterostructure.  We have developed
a novel, two-color, time-resolved magneto-optical Kerr effect (MOKE)
spectroscopy setup, which allows us to create transient carriers
{\em only in the magnetic InMnAs layer} using mid-infrared (MIR) pulses and
then probe the induced magnetization changes through the MOKE angle of
near-infrared (NIR) probe pulses.  Our data indeed shows that magnetic
properties, e.g., coercivity and remanent magnetization, can be significantly
modified by intense MIR pulses.  We performed simultaneous measurements
of MOKE angle and reflectivity for examining spin and charge dynamics
separately.  Furthermore, coherent optical spin injection using different
senses of circular polarization led to different signs for the net MOKE
changes induced by the pump.
%Finally, the
%temperature and pump intensity dependences of these effects will be
%shown and discussed.

%-----------------------------------------------------------------------
\section{Experimental Details}
%-----------------------------------------------------------------------

We performed two-color time-resolved MOKE spectroscopy using femtosecond
pulses of MIR and NIR radiation.  The experimental setup is schematically
shown in Figure 1.  The source of intense MIR pulses was an optical
parametric amplifier (OPA) (Model FS-TOPAS-4/800,
Quantronix/Light Conversion) pumped by a Ti:Sapphire-based regenerative
amplifier (Model CPA-2010, Clark-MXR, Inc., 7300 West Huron River Drive,
Dexter, MI 48130).  The OPA was able to produce tunable and intense radiation
from 522 nm to 20 $\mu$m using different mixing crystals.  The CPA
produced pulses of NIR radiation with a wavelength of 775 nm, a pulse
energy of $\sim$ 1 mJ, and a pulse duration of $\sim$ 140 fs at a tunable
repetition rate of 50 Hz $-$ 1 kHz.  We used a very small fraction
($\sim 10^{-5}$) of the CPA beam as a probe and the output beam from
the OPA tuned to 2 $\mu$m as the pump.  The CPA probe beam went through a
computer-controlled variable delay stage in addition to a fixed $\sim$ 2 m
long delay stage that equalized the pump and probe beam path lengths to the
sample by taking into account the total path length inside of the
multi-pass OPA.  The two beams were made collinear by a non-polarizing
beam splitter (Lambda Research Optics, Inc.), which was 50\% reflective to
the NIR probe and 50\% transmissive to the MIR pump, and then focused by
an off-axis parabolic mirror with a six inch focal length onto the sample
mounted inside a 10 Tesla superconducting magnet with ZnS cold windows
and CaF$_2$ room temperature windows.   The reflected NIR probe beam entered
a Wollaston prism which spatially separated the $s$- and $p$-components of
the probe beam, which we then focused on a balanced bridge system consisting
of a Si detector pair.  The signal from the balanced detectors,
which was proportional to the induced MOKE angle change, was fed into
a lock-in amplifier or a boxcar integrator and was recorded by a computer.
An additional beam splitter was placed before the Wollaston prism for
monitoring the reflectivity of the probe.  With this setup, we were able to
record the MOKE angle and reflectivity as functions of time delay
and magnetic field.

At this pump wavelength (2 $\mu$m), the photon energy (0.62 eV) was smaller
than the band gaps of the GaSb buffer (0.812 eV) and GaAs substrate
(1.519 eV) but larger than that of InMnAs ($\sim$ 0.42 eV), so the pump
created carriers only in the InMnAs layer.  The beam diameter of the MIR
pump at the sample position was measured by pinholes to be $\sim$ 75 $\mu$m.
The maximum MIR pulse energy used in this experiment was $\sim$ 6.3 $\mu$J,
which corresponds to a fluence of $\sim$ 0.45 J/cm$^2$.  Using the optical
constants of InAs and the thickness of the InMnAs layer, we estimated the
maximum density of photocreated carriers to be $\sim$ 7.5
$\times$ 10$^{22}$ cm$^{-3}$, which is an extremely large number,
especially if we think of the fact that the density of Mn ions is
only $\sim$ 10$^{21}$ cm$^{-3}$.
 
The sample studied was an InMnAs/GaSb single heterostructure with a Curie
temperature ($T_c$) of 55 K, consisting of a 25 nm thick
In$_{0.91}$Mn$_{0.09}$As magnetic layer and an 820 nm thick GaSb buffer layer
grown on a semi-insulating GaAs (100) substrate.  Its room temperature
hole density and mobility were 1.1 $\times$ 10$^{19}$ cm$^{-3}$ and
323 cm$^2$/Vs, respectively, estimated from Hall measurements.
The sample was grown by low temperature molecular beam epitaxy (growth
conditions described previously\cite{tom}) and then annealed at
250 $^{\circ}$C, which increased the $T_c$ by $\sim$ 10 K.
\cite{hayashi,potashnik}  The magnetization easy axis was perpendicular
to the epilayer due to the strain-induced structural anisotropy caused
by the lattice mismatch between InMnAs and GaSb (InMnAs was under tensile
strain).  This allowed us to observe ferromagnetic hysteresis loops in
the polar Kerr configuration.

%-----------------------------------------------------------------------
\section{Experimental Results and Discussion}
%-----------------------------------------------------------------------

Figure 2(a) shows three magnetic-field-scan data exhibiting ferromagnetic
hysteresis loops at a temperature of 16 K, taken under different conditions.
The curve labeled 'No Pump' was taken with the OPA pump beam blocked,
while the other two curves were taken under high OPA excitation (fluence
$\sim$ 0.2 J/cm$^2$) with a time delay of 0 ps and $-$7 ps, respectively.
It can be seen that at timing zero the loop {\em horizontally}
collapsed, i.e., the coercivity is almost zero.  Note that this curve
is not intentionally offset; namely, this vertical shift is a real effect
induced by the pump, which exists only for a short time ($\sim$ 2 ps).
As discussed later, we attribute this transient vertical shift of the MOKE
signal to the coherent spin polarization of the photo-generated carriers.
The negative time delay data also shows similar horizontal
shrinkage though it is not as dramatic as the timing zero data.
It is important to note that the vertical height (i.e., remanence) of
the loops is not much affected by the pump, which excludes simple lattice
heating as the cause of the horizontal loop quenching.  To support this view
more convincingly, we show in Fig. 1(b) CW magnetic circular dichroism
data taken for the same sample at 10 K, 35 K, 45 K and 55 K.  As can been seen
clearly, raising the lattice temperature results in dramatic loop
shrinkage {\em both horizontally and vertically}.  Furthermore, we took
time-resolved MOKE data with various combinations of average powers
and fluences (data not shown) and found that {\em it is the fluence that
determines the degree of collapse, not the average power}.  For example,
data taken at a low repetition rate with a high fluence displayed
significant quenching while data taken at a high repetition
rate with a low fluence did not show any quenching.

Next we discuss some time-scan data.  Figure 3(a) shows two traces
representing the pump-induced MOKE signal change versus time delay taken
under the excitation of the MIR pump with two opposite senses of
circular polarization, i.e., $\sigma^+$ and $\sigma^-$.  As clearly
demonstrated here, opposite polarizations result in {\em opposite signs}
of the photo-induced MOKE change, suggesting that these fast decaying
transient MOKE signals are due to the photo-induced coherent carrier spin
polarization.  The pump-induced reflectivity change as a function of time
is shown in Fig. 3(b).  One can see a quick disappearance of the
induced reflectivity change.  In metals, photo-induced transmission or
reflection changes can be attributed to electron temperature changes,
which show characteristic cooling behavior as the electron system
loses it energy to the lattice.  In the present semiconductor system,
however, the reflectivity change is most likely due to the transient
change of the carrier density.  The observed fast (< 2 ps) decay is
consistent with the fast decays typically observed in low temperature
grown semiconductors such as those used in terahertz emitters and
receivers.\cite{smith}  To further understand the origin of the
polarization-dependent ultrafast MOKE change, we did the same
measurements at elevated temperatures.  As an example, data taken at
122 K is shown in Fig. 4.  Here, again, we see opposite signs for
$\sigma^+$ and $\sigma^-$ polarization.  We attempted to observe photo-induced
ferromagnetism by scanning the magnetic field at these high temperatures
but did not see any evidence.  These facts lead us to believe that these
signals are related to the coherent carrier spins.
We do not have an explanation for the second peak around 1 ps observed for
both polarizations.  More detailed temperature dependent measurements
are in progress to elucidate this feature.

%Figure 4 shows a time-scan trace [in (a)] together with three
%magnetic-field-scan data [in (b)-(d)] taken at different time delays.  Here
%the pump fluence was kept low enough to suppress any noticeable
%horizontal loop shrinkage (fluence $\sim$ 0.1 J/cm$^2$).  Even at this
%low fluence, the pump-induced coherent carrier spin signal exists [see
%Fig. 4(a)].  Under these conditions, we carefully compared the vertical
%height of the hysteresis loops, which is proportional to the remanent
%magnetization, in order to deduce the time evolution of the induced
%magnetization change and to see there is any interplay between carrier
%spin dynamics and localized magnetization dynamics, and the result is
%shown in the inset of Fig. 4(a).  We can observe a small but definite
%decrease in loop height, which is most significant at timing zero and
%decays within a xx ps.  More experiments are underway to further
%explore the physics of photo-induced ultrafast spin and magnetization
%dynamics in ferromagnetic semiconductors.

%-----------------------------------------------------------------------
\section{Summary}
%-----------------------------------------------------------------------

We have presented results of what we believe to be the first ultrafast optical
study of spin/magnetization dynamics in ferromagnetic InMnAs/GaSb
heterostructures.  Using a novel two-color time-resolved magneto-optical
Kerr effect spectroscopy technique, we created transient carriers
only within the ferromagnetic InMnAs layer using intense
mid-infrared pulses and observed the induced magnetization changes
through the Kerr angle of near-infrared probe pulses.  Our data shows
that magnetic properties, particularly coercivity,
can be drastically modified by the intense MIR pulses.  We were able to
study spin and charge dynamics separately by simultaneously measuring MOKE
and reflectivity.  Finally, coherent optical spin injection using
different senses of circular polarization led to different signs for the
net MOKE changes induced by the pump.

%%%%%%%%%%%%%%%%%%%%%%%%%%%%%%%%%%%%%%%%%%%%%%%%%%%%%%%%%%%%%%%%%%%%%%%%
%                          Acknowledgements
%%%%%%%%%%%%%%%%%%%%%%%%%%%%%%%%%%%%%%%%%%%%%%%%%%%%%%%%%%%%%%%%%%%%%%%%
\begin{acknowledgments}
This work was supported by the Defense Advanced Research Projects Agency
through grant No. MDA972-00-1-0034 and the National Science Foundation
through grant No. DMR-0134058 (CAREER).  We are grateful to Bruce Brinson
and Ben Schmid for their technical assistance.
\end{acknowledgments}

%%%%%%%%%%%%%%%%%%%%%%%%%%%%%%%%%%%%%%%%%%%%%%%%%%%%%%%%%%%%%%%%%%%%%%%%
%                          References
%%%%%%%%%%%%%%%%%%%%%%%%%%%%%%%%%%%%%%%%%%%%%%%%%%%%%%%%%%%%%%%%%%%%%%%%

%\section{References}
%\bibliography{paper}

%\newpage

\newpage
%\twocolumngrid

%.......................................................................
%                              Figure 1
%.......................................................................
\begin{figure}
\includegraphics[scale=0.6]{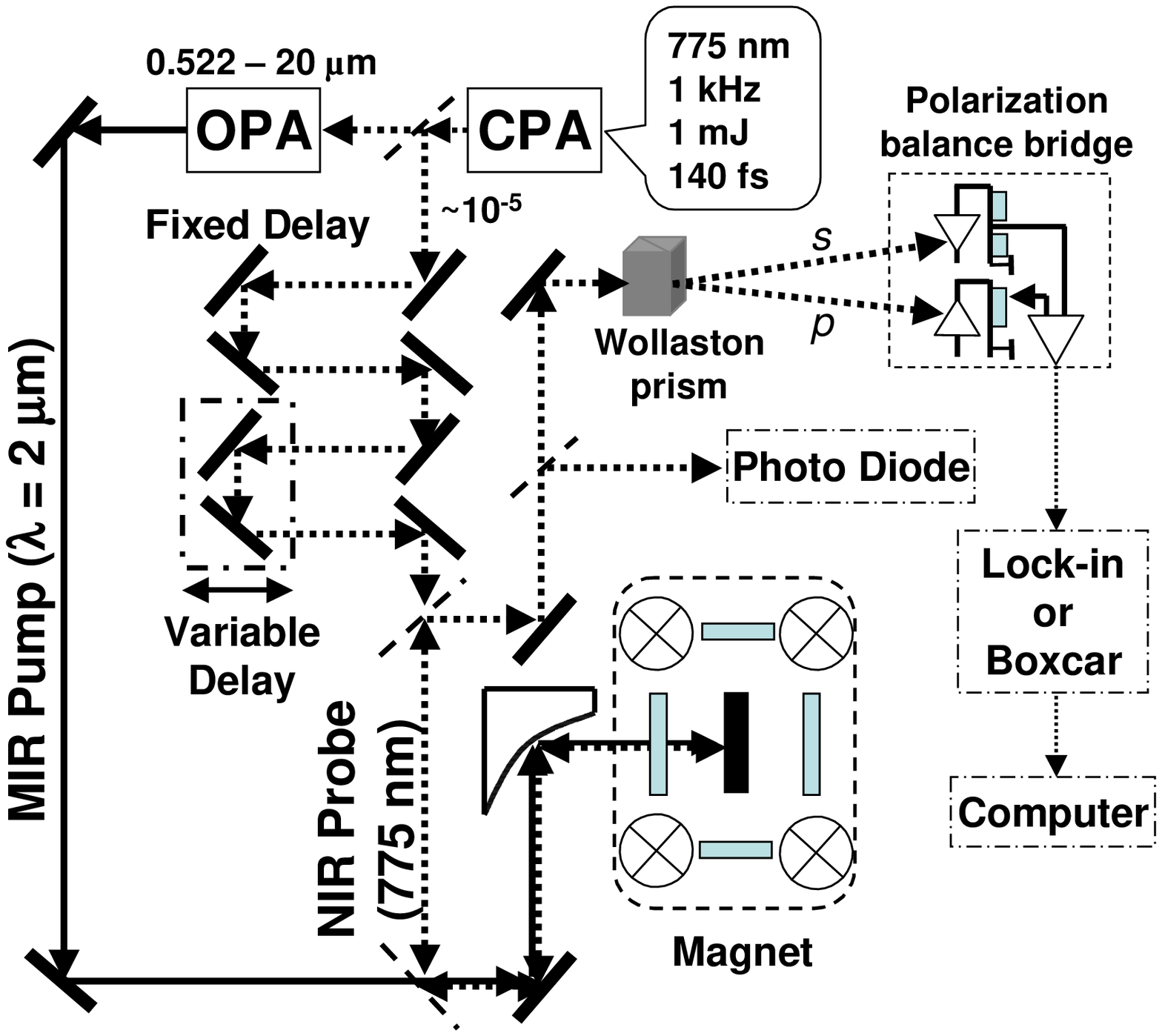}
\caption {Schematic diagram for the two-color time-resolved MOKE
spectroscopy setup.  OPA: optical parametric amplifier.  CPA: chirped pulse
amplifier.  The CPA is a Ti:Sapphire-based regenerative amplifier
(Model CPA-2010, Clark-MXR, Inc.).  A small fraction
($\sim 10^{-5}$) of the CPA beam is used as a probe and the OPA beam
tuned to 2 $\mu$m is used as the pump.}
\label{fig1}
\end{figure}
%.......................................................................

\newpage
%.......................................................................
%                              Figure 2
%.......................................................................
\begin{figure}
\includegraphics[scale=0.74]{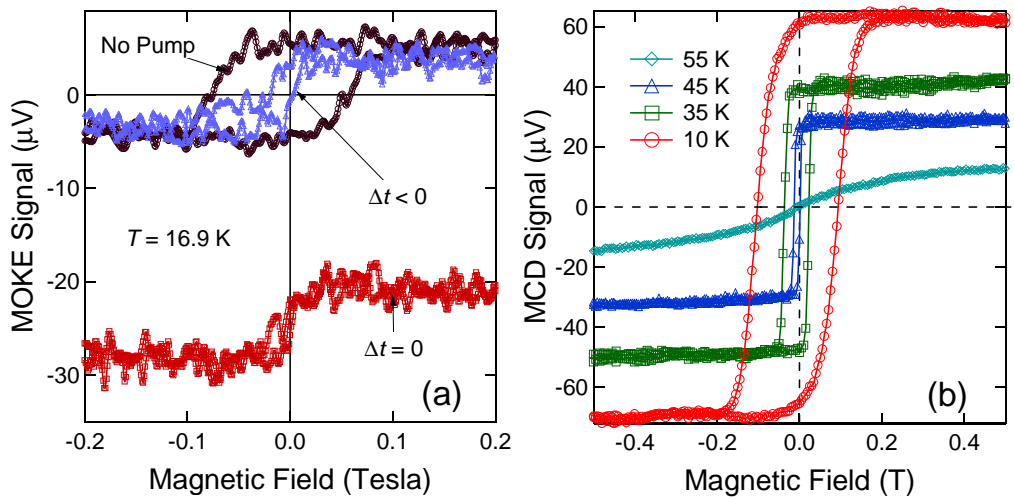}
\caption{(a) Black circles: MOKE signal versus magnetic field with
no pump.  Red squares: MOKE signal versus magnetic fields under MIR pump
excitation ($\sim$ 0.2 J/cm$^2$) at timing zero.  Blue triangles:
MOKE signal versus magnetic field under MIR pump excitation ($\sim$
0.2 J/cm$^2$) at a time delay of $-$7 ps.
(b) CW magnetic circular dichroism data taken at four different
temperatures (10 K, 35 K, 45 K, and 55 K).
}
\label{fig2}
\end{figure}

\newpage
%.......................................................................
%                              Figure 3
%.......................................................................
\begin{figure}
\includegraphics[scale=0.72]{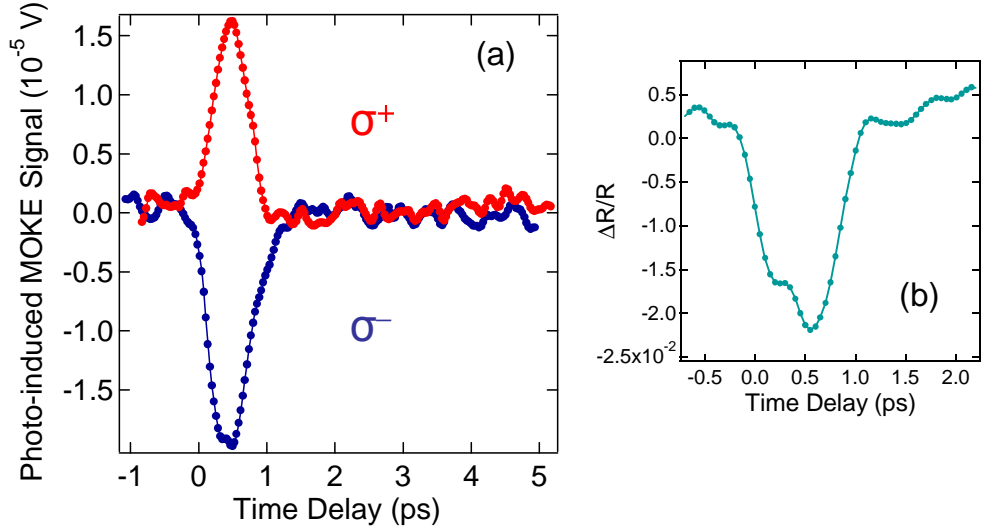}
\caption{(a) Photo-induced MOKE signal at a temperature of 16 K
versus time delay under pumping with circularly polarized MIR radiation.
(b) The reflectivity of the NIR probe is plotted as a function of
time delay, showing fast carrier recombination.}
\label{fig3}
\end{figure}

\newpage
%.......................................................................
%                              Figure 4
%.......................................................................
\begin{figure}
\includegraphics[scale=0.8]{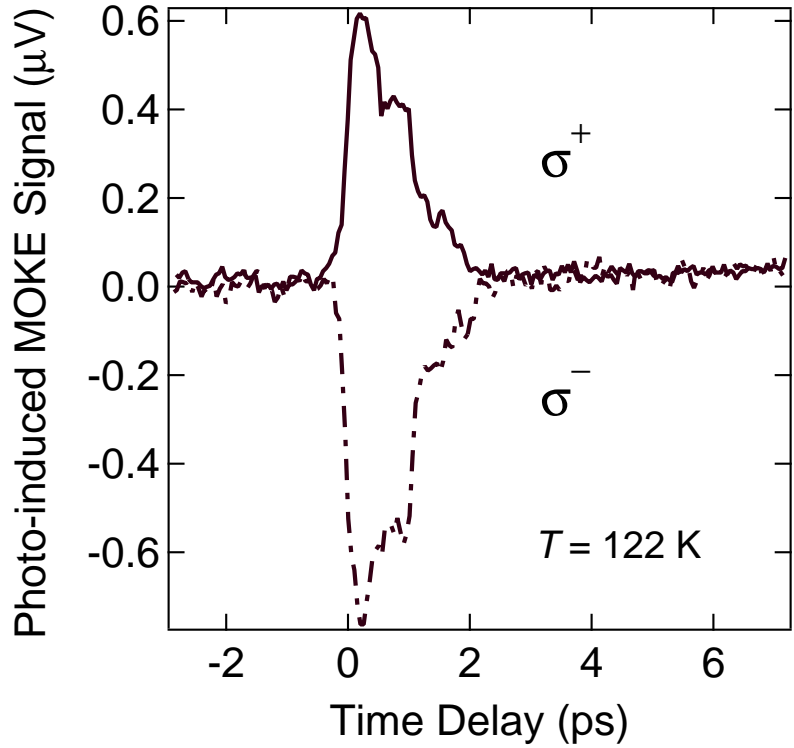}
\caption{Photo-induced MOKE signal versus time delay for an
InMnAs/GaSb heterostructure with a Curie temperature of 55 K, taken at
122 K under pumping with circularly polarized MIR radiation.
}
\label{fig3}
\end{figure}
%.......................................................................

%\onecolumngrid
\end{document}